\documentclass[12pt,preprint]{aastex}
\usepackage{rotating}
\slugcomment{}

\shorttitle{Quasi--Circular Ribbon Flare}
\shortauthors{Joshi et al.}

\begin{document}

%------------------------------------------------------------------------------------------

\title{The Role of Erupting Sigmoid in Triggering a Flare with Parallel and Large--Scale Quasi--Circular Ribbons}

%------------------------------------------------------------------------------------------

\author{Navin Chandra Joshi\altaffilmark{1}}
	\affil{School of Space Research, Kyung Hee University, Yongin, Gyeonggi--Do, 446--701, Korea; navin@khu.ac.kr, njoshi98@gmail.com}
\author{Chang Liu\altaffilmark{2,3}}
	\affil{Space Weather Research Laboratory, New Jersey Institute of Technology, University Heights, Newark, NJ 07102-1982, USA}
	\affil{Big Bear Solar Observatory,  40386 North Shore Lane, Big Bear City, CA 92314, USA}
\author{Xudong Sun\altaffilmark{4}}
	\affil{W.W.Hansen Experimental Physics laboratory, Stanford University, Stanford, CA 94305, USA}
\author{Haimin Wang\altaffilmark{2,3}}
	\affil{Space Weather Research Laboratory, New Jersey Institute of Technology, University Heights, Newark, NJ 07102-1982, USA}
	\affil{Big Bear Solar Observatory,  40386 North Shore Lane, Big Bear City, CA 92314, USA}
\author{Tetsuya Magara\altaffilmark{1,5}}
	\affil{School of Space Research, Kyung Hee University, Yongin, Gyeonggi--Do, 446--701, Korea; navin@khu.ac.kr, njoshi98@gmail.com}
	\affil{Department of Astronomy and Space Science, Kyung Hee University, Yongin, Gyeonggi-Do, 446-701, Korea}
\author{Y.-J. Moon\altaffilmark{1,5}}
	\affil{School of Space Research, Kyung Hee University, Yongin, Gyeonggi--Do, 446--701, Korea; navin@khu.ac.kr, njoshi98@gmail.com}
	\affil{Department of Astronomy and Space Science, Kyung Hee University, Yongin, Gyeonggi--Do, 446--701, Korea}
%------------------------------------------------------------------------------------------

\altaffiltext{1}{School of Space Research, Kyung Hee University, Yongin, Gyeonggi--Do, 446--701, Korea; navin@khu.ac.kr, njoshi98@gmail.com}
\altaffiltext{2}{Space Weather Research Laboratory, New Jersey Institute of Technology, University Heights, Newark, NJ 07102--1982, USA}
\altaffiltext{3}{Big Bear Solar Observatory,  40386 North Shore Lane, Big Bear City, CA 92314, USA}
\altaffiltext{4}{W.W.Hansen Experimental Physics laboratory, Stanford University, Stanford, CA 94305, USA}
\altaffiltext{5}{Department of Astronomy and Space Science, Kyung Hee University, Yongin, Gyeonggi--Do, 446--701, Korea}
%------------------------------------------------------------------------------------------

\begin{abstract}

In this paper, we present observations and analysis of an interesting sigmoid formation, eruption and the associated flare that occurred on 2014 April 18 using multi--wavelength data sets. We discuss the possible role of the sigmoid eruption in triggering the flare, which consists of two different set of ribbons: parallel ribbons as well as a large--scale quasi--circular ribbon. Several observational evidence and nonlinear force--free field extrapolation results show the existence of a large--scale fan--spine type magnetic configuration with a sigmoid lying under a section of the fan dome. The event can be explained with the following two phases. During the pre--flare phase, we observed the formation and appearance of sigmoid via tether--cutting reconnection between the two sets of sheared fields under the fan dome. The second, main flare phase, features the eruption of the sigmoid, the subsequent flare with parallel ribbons, and a quasi--circular ribbon. We propose the following multi--stage successive reconnections scenario for the main flare. First, tether--cutting reconnection is responsible for the formation and the eruption of the sigmoid structure. Second, the reconnection occurred in the wake of the erupting sigmoid produces the parallel flare ribbons on the both sides of the circular polarity inversion line. Third, the null--type reconnection higher in the corona, possibly triggered by the erupting sigmoid, leads to the formation of a large quasi--circular ribbon. For the first time we suggest a mechanism for this type of flare consisting of a double set of ribbons triggered by an erupting sigmoid in a large scale fan--spine type magnetic configuration.

\end{abstract}

%------------------------------------------------------------------------------------------

\keywords{Sun: activity -- Sun: flares -- Sun: magnetic fields -- Sun: X--rays, gamma rays}

%------------------------------------------------------------------------------------------

\section{Introduction}
\label{}
Solar flares are the phenomena characterized as the release of free magnetic energy due to magnetic reconnection in the corona. The released magnetic energy is converted into kinetic energy to accelerate electrons and other energetic particles as well as thermal energy to heat the plasma \citep[and references cited therein]{Benz08,Schrijver09,Shibata11,Fletcher11}. Formation of solar flare ribbons on both sides of the magnetic polarity inversion line (PIL) is known as a secondary process, when accelerated electrons hit the low solar atmosphere \citep{Fletcher01,Krucker11}. Many models and theories have been proposed to interpret these energetic events on the Sun since the first white--light solar flare was observed by R.C. Carrington and R. Hodgson in 1859 \citep{Carrington59,Hodgson59}. 

The ``CSHKP" model based on the work of \cite{Carmichael64}, \cite{Sturrock66}, \cite{Hirayama74}, and \cite{Koop76} is a well--accepted two--dimensional (2D) model for solar flares \citep[and references therein]{Shibata98}. Solar flare events have been interpreted in terms of the classical CSHKP model \citep{Magara96,Joshi13}. Apart from this 2D model, three--dimensional (3D) reconnection models for triggering solar flares have been also discussed e.g., 3D extension of the standard flare model \citep{Aulanier12}, the fan--spine 3D magnetic reconnection model \citep{Lau90,Shibata94,Aulanier00,Torok09,Guglielmino10,Liu11,Deng13}, and the slipping reconnection model \citep{Aulanier06,Dudik14}. Two--ribbon flares are usually caused by the reconnection among the surrounding arcades underneath the erupting filaments or sigmoids \citep{Liu07,Aulanier10}, while circular ribbon flares are due to null--point reconnection in a complex fan--spine magnetic configuration. Evidence of circular ribbon flares have been presented by several authors \citep{Masson09,Su09,Reid12,Wang12,Sun13,Deng13,Jiang13,Liu13,Jiang14,Vemareddy14b,Yang15}. \cite{Masson09} used {\it Transition Region and Coronal Explorer} (TRACE) and {\it Solar and Heliospheric Observatory/Michelson Doppler Imager} (SOHO/MDI) observations to study the nature of a circular ribbon flare that occurred on 2002 November 16 and the associated magnetic topology. \cite{Su09} studied a B1.7 class solar flare in a decaying active region (AR) with a quasi--circular ribbon on 2008 May 17. \cite{Reid12} also explained the evolution of 2002 November 16 solar flare using {\it Reuven Ramaty High Energy Solar Spectroscopic Imager} (RHESSI) X--ray sources and TRACE UV observations. \cite{Wang12} analyzed five circular ribbon flares that were associated with surges using blue--wing H$\alpha$ observations from {\it Big Bear Solar Observatory} on 1991 March 17--18. \cite{Deng13} presented high--resolution spectroscopic H$\alpha$ imaging observations of a circular ribbon flare on 2011 October 22 using data from {\it Interferometric Bidimensional Spectrometer}. \cite{Sun13} analyzed and found the signature of hot spine loops joining the quasi--circular ribbon and the remote brightening on 2011 November 15. \cite{Jiang13} and \cite{Jiang14} studied the formation and eruption of an AR sigmoid from below the fan dome and the triggering of a circular ribbon flare on 2011 September 6. \cite{Liu13} presented the $He_{I}~D3$ observation of the M6.3 class solar flare on 1984 May 22 and found the signature of a huge circular ribbon. Recently, \cite{Vemareddy14b} studied the quasi--elliptical X1.5 solar flare that occurred on 2011 March 9 with two remote ribbons. More recently, \cite{Yang15} also investigated an X--class circular ribbon solar flare that occurred on 2012 October 23. All these observational studies of circular ribbon flares suggest a fan--spine type magnetic topology and the reconnection at a coronal null point \citep{Torok09,Sun13}. This type of reconnection may occur between the inner and outer field lines in a fan--dome magnetic configuration separated by separatrix layers \citep{Pariat09}. Along with these observational works, there are a few attempts to simulate these kind of circular ribbon flares \citep[e.g.,][]{Masson09}.

Sigmoidal structures are the observational signature of the highly twisted and sheared magnetic fields in the solar corona. Sigmoidal shapes in the soft X--ray as well as in EUV emissions provide the storage of free magnetic energy in the solar corona and are the best proxies for the eruptive filaments, prominences, and flux ropes \citep{Gibson02,Green07,Savcheva14}. Various observational studies as well as simulations of these structures have been carried out in order to understand their formation, eruption, and the associated solar flares \citep{Sterling00,Savcheva12,Schmieder13,Jiang13,Vemareddy14a,Jiang14,Joshi14a}. Photospheric magnetic cancellation \citep{van89} and the coronal tether--cutting reconnection \citep{Sturrock84,Moore01} are two main mechanisms known for the buildup as well as the initial rising of sigmoid structures in the solar corona. Several studies have been performed showing the formation of sigmoid with these mechanisms \citep{Inoue11,Liu12,Liu13,Joshi14b,Vemareddy14a}. Other mechanisms such as torus instability \citep{Kliem06}, kink instability \citep{Torok04}, and the magnetic breakout \citep{Antiochos99} are well known to drive the eruptions. Torus instability occurs when the external field decreases very fast with the coronal heights (i.e., the decay index $>$1.5) \citep{Aulanier10,Demoulin10}. Kink instability occurs when the twist of the field lines about the flux rope axis exceeds a critical value of about 1.5--2 \citep{Torok05,Srivastava10}. In the breakout reconnection, the removal of overlying field lines due to reconnection at the top of the core field lines drives the eruption \citep{Sterling04,Karpen12}. 

Several attempts have been in progress to understand the role of the erupting sigmoids and filaments from underneath the fan--dome structures in triggering solar flares with two different set of flare ribbons: parallel ribbons as well as circular ribbons. In this work, we present the observations of such a solar flare producing two parallel ribbons as well as a large quasi--circular ribbon triggered by an erupting sigmoid in a complex and large--scale fan--spine type magnetic configuration. We interpret the main flare in the context of different stages of reconnections which occurred during the sigmoid eruption. The paper is structured as follows: Section 2 presents the observational data sets and the multiwavelength observational analyses are presented in Section 3. Nonlinear force--free field (NLFFF) extrapolations are described in Section 4. In Section 5, the main results and their discussions are described.

%------------------------------------------------------------------------------------------

\section{Observational Data Sets}
\label{}

For the present work we use data from several instruments. Data in ultra--violet (UV) and extreme ultra--violet (EUV) wavelengths are taken from {\it Solar Dynamics Observatory/Atmospheric Imaging Assembly} \cite[SDO/AIA;][]{Lem12}. Magnetic field data is collected from {\it SDO/Helioseismic Magnetic Imager} \citep[SDO/HMI;][]{Schou12,Hoeksema14}. Both AIA and HMI are pay--load instruments on SDO. The pixel sizes of AIA and HMI are around 0.6\arcsec~and 0.5\arcsec, with time cadences of 12 s and 45 s, respectively. {\it Reuven Ramaty High Energy Solar Spectroscopic Imager} (RHESSI; \cite{Lin02}) is an X--ray imager observing the Sun in different energy bands between 3--1500 keV. We construct X--ray images using the PIXON algorithm with an integration time of 20 s.

%------------------------------------------------------------------------------------------
\section{Observational Analysis}
\label{}
%------------------------------------------------------------------------------------------
\subsection{Active Region Location and Temporal Profiles}
\label{}

Figure 1(a) presents a SDO/AIA 304 \AA~image, showing the disk view of the Sun on 2014 April 18 at 11:30:07 UT. The dotted curved lines over the Sun represent the grid lines of $10^{\circ}$ and the limb is shown by the solid black line. The over plotted black box in the southeast part of the Sun shows the region under study. Figures 1(b) and 1(c) show the zoomed--in views of the line--of--sight magnetogram and the white--light image, respectively at 11:30 UT corresponding to the black box shown in Figure 1(a). It can be seen that the activity region lies on the southwest quadrant ($\approx{S18 W37}$) of the solar disk. As a matter of fact, there are three ARs adjacent to one another in this area. They are NOAA ARs 12036, 12037, and 12043, as represented by the arrows in Figures 1(b) and 1(c). The two main ARs 12036 and 12037 consist of a $\beta\gamma$ type magnetic configuration, while the AR 12043 has an $\alpha$ type magnetic configuration on that day. Another small AR, i.e., AR 12041 is also seen in the southwest of the main AR 12036. Because of their small size, the ARs 12043 and 12041 are not visible in the SDO/AIA continuum image (Figure 1(c)). It is evident that the area includes three ARs is highly complex as far as the photospheric magnetic structure is concerned. 

Figure 2(a) shows the {\it Geostationary Operational Environmental Satellite} (GOES) and RHESSI X--ray flux variation with time between 11:00 UT to 15:00 UT on 2014 April 18. The solid and dotted black curves show the GOES X--ray flux profiles in 1.0--8.0 and 0.5--4.0 \AA, respectively. The temporal profiles of RHESSI X--ray flux in different energy ranges are overplotted with different colors, i.e., 3--6 (pink), 6--12 (red), 12--25 (green), 25--50 (blue), and 50--100 (yellow) keV. We observe three peaks at about 11:50 UT, 12:35 UT and 13:00 UT, corresponding to three different stages of energy release, respectively (i.e., the preflare, nearby jet activity, and the main flare). These peaks are observed in almost all the light curves and are marked by arrows. The preflare activity appeared in the GOES and RHESSI X--ray profiles between about 11:40 and 12:10 UT. The main flare phase started at about 12:31 UT and peaked at about 13:03 UT. The decay phase is very long and last till about 14:40 UT. The main M7.3 class flare can be considered as a long duration flare event. The RHESSI X--ray flux observations are available only during the late impulsive and early decay phases of the main flare phase from about 12:46 UT to 13:13 UT. A small peak corresponding to the jet--activity is observed between about 12:34 and 12:36 UT in the GOES time profile. No RHESSI flux observations are available at this phase. The small peak corresponding to the jet may be a combination of X--ray flux coming from the jet activity and the tether--cutting reconnection at the start of the main phase.

In order to check the spatial location of these peaks on the solar disk, we plot the SDO/AIA intensity curves at different wavelength bands between 11:30 UT and 12:46 UT (Figure  2(b)). The region used for the estimation of the intensity light curves is shown by the white box in Figure 1(a), covering the region of pre and main flares including the jet. We calculate the average count values of the flaring region and divide them by the average background counts. All the curves are normalized in order to make a better comparison. It can be seen that the intensity variation at hotter channels are similar to that of the integrated GOES curves. We find that the contribution of the active region flux alone can account for most of the full--disk variation. It can be concluded that all the three peaks are from our flaring region i.e., from AR 12036.
%------------------------------------------------------------------------------------------

\subsection{Preflare Activity Phase: Formation and Appearance of Sigmoid}
\label{}

The overall evolution of the event can be described by two different phases. The first phase is the preflare phase (sigmoid formation), followed by the main flare phase, which includes the nearby jet--activity phase. Figures 3(a)--(e) show the sequence of the selected SDO/AIA 131 \AA~wavelength images which demonstrate the preflare activity phase from 11:35 UT to 12:10 UT. SDO/AIA 131 \AA~images provide information about hotter flaring regions in the solar corona. Figure 3(f) presents a SDO/HMI line--of--sight magnetogram at 12:10:19 UT. The overplotted letters P and N represent the locations of positive and negative polarity regions at the main AR, respectively. Mainly four patches of polarities i.e., two positive (P1 and P2) and two negative (N1 and N2) are observed. The curved PIL lies between these negative and positive polarity regions.

The preflare phase started with a compact brightening at 11:35 UT (Figure 3(a)). This brightening is most probably due to the interaction and reconnection between the northern and southern sheared field lines. The northern sheared field lines connect the regions P1 and N1 while the southern field lines connect the regions N2 and P2. The reconnection is evidenced by the observed RHESSI 3--6 and 6--12 keV sources over the reconnection region (Figures 3(b) and 3(c)). All the RHESSI sources are mostly in the soft X--ray (SXR) range and they should lie in the corona. We do not observe any RHESSI hard X--ray (HXR) footpoint sources, possibly because it is a weak event. The reconnection continued at the PIL between N1 and P2 from 11:45 UT to 12:10 UT, resulting in the formation and appearance of a large sigmoid (Figures 3(d) and 3(e)). The filling of hot plasma in the whole structure makes it visible. The full appearance of the sigmoid can be seen at 12:10:08 UT (Figure 3(e)). A sketch diagram of the sigmoid based on the SDO/AIA 131 \AA~images is drawn over the SDO/HMI image in Figure 3(f). It is evident from this image that the southern/northern footpoints of the formed sigmoid lie in the negative (N2)/positive (P1) polarity regions, respectively. It is also observed that after its formation, the sigmoid remains at some height in the corona until $\approx$12:30 UT. It also becomes invisible soon after its full appearance in hotter AIA channels. The dynamics of the sigmoid formation is clear in the AIA 94 and 131 \AA~movies (see animations associated with Figure 3). A schematic picture of the preflare scenario is represented in Figure 4 for better understanding in a simple way. The sigmoid lies in the southern hemisphere and it is S--shaped, which is consistent with the sigmoid hemispheric rule \citep[c.f.,][]{Savcheva14}.
%------------------------------------------------------------------------------------------

\subsection{Main Flare Phase: Sigmoid Eruption and M7.3 Solar Flare}
\label{}

The main flare phase started at $\approx$12:31 UT followed by the nearby jet activity phase started at $\approx$12:34 UT. Figures 5(a)--(f) show the sequence of selected SDO/AIA 131 \AA~images from 12:31 to 13:00 UT, showing the observations of the main flare phase and jet--activity phase. Figures 5(g)--(i) present the SDO/AIA 171 \AA~images during the decay phase of the main flare from 13:05 to 13:45 UT. The main flare phase started with a compact brightening near the PIL between polarities N1 and P2 (Figure 5(a)). This may be due to the tether--cutting type reconnection at the middle of the sigmoid between the sheared field lines. A few minutes later, we found a jet--like ejection near the compact brightening area from 12:34 UT to 12:36 UT (Figures 5(a) and (b)). Figures 6(a)--(e) show the zoomed--in view of the jet in five SDO/AIA wavelength channels. The selected region for the zoomed--in view is shown in Figure 5(b) by the white box. The appearance of the jet in different channels shows that the jet contained multi--thermal plasma. Figure 6(f) shows the SDO/HMI line--of--sight magnetogram overplotted by the SDO/AIA 304 \AA~intensity contours. Apparently, the base of the jet lies at the small negative polarity region, pointed by the green arrow in Figure 6(f). 

The initial slow eruption of the sigmoid started simultaneously with the main flare phase at 12:31 UT. The low--lying tether--cutting type reconnection may be responsible for the eruption of the sigmoid. The quantitative measurements of the projected height of the jet plasma ejection and the sigmoid eruption are shown in Figure 7. The left panel of Figure 7 shows the SDO/AIA 94 \AA~image at 12:39:01 UT, with overplotted trajectories along which the projected heights are measured. The bottom most point of each trajectory is used as the reference point for height measurements. The height--time measurements of the jet plasma ejection (green curve) and the sigmoid eruption (red curve) are presented in Figure 7(b). These measurements are made using SDO/AIA 94 \AA~images with a temporal cadence of 12 s. The average ejection speed of the jet plasma is about 122 $\rm km~s^{-1}$ between 12:33 UT and 12:40 UT. The sigmoid height--time profile shows two different stages of eruptions. The initial slow speed comes out to be about 10 $\rm km~s^{-1}$ between 12:30 UT and 12:37 UT. Then the motion of the sigmoid is accelerated to a higher speed of about 45 $\rm km~s^{-1}$ between $\approx$12:37 UT and $\approx$12:44 UT. The speeds are calculated from the linear fit to the measured height--time data. In order to understand the relation between the sigmoid eruption and the flare, the GOES X--ray fluxes in 1--8 and 0.5--4 \AA~channels are overplotted. It is evident that the GOES flux enhancement is simultaneous with the sigmoid eruption. The overall dynamics of the sigmoid eruption can be seen in the SDO/AIA 131 \AA~running difference images (Figure 8 and the associated animation).

It is evident from the observations that the main flare started with the eruption of sigmoid and formed extended flare ribbons (Figures 5(c)--(f) and the associated animation). Figure 9 shows SDO/AIA 1600 \AA~images presenting the evolution of flare ribbons. Two sets of flare ribbons are observed; the first set includes the two parallel ribbons on both sides of the inner curved PIL, and the second is the large--scale circular ribbon. The two parallel ribbons appeared at 12:35 UT (marked as R1 and R2 in Figure 9(b)) about 4 minutes later than the start of the sigmoid slow rising motion. Afterward, the size as well as the brightening of both the ribbons increased simultaneously with the sigmoid eruption (Figures 9(b)--(d)). The well--developed parallel ribbons appeared at around 12:44 UT (Figure 9(d)). At this time the sigmoid rose to a height of about 2.6 Mm from its original height of 3.2 Mm. A continuous separation between the parallel ribbons has also been observed (see the animations associated with Figure 9). The brightening of the circular ribbon, which is located to the northeast of both the parallel ribbons, started at 12:42 UT (Figure 9(d)), about 11 minutes after the start of the sigmoid slow eruption. At this time the approximate height of sigmoid is about 4.86 Mm. The brightening in the western part of the circular ribbon started at 12:45 UT (Figure 9(e)). Thereafter the brightening extends eastward up to 12:50 UT (Figure 9(f)). However, we do not find much outward expansion of the circular ribbon. All the dynamical evolution of the ribbon formation and brightening extension can be clearly seen in the SDO/AIA 1600 \AA~movie (see the animation associated with Figure 9). Figures 10(a)--(e) shows the well--developed circular flare ribbon as well as the two parallel ribbons in five EUV wavelength channels of AIA at 12:50 UT. The circular ribbon is marked by the white arrows. Figure 10(f) shows the SDO/HMI image overplotted by the AIA 1600 \AA~image intensity contours in red color. The two parallel ribbons are formed after the reconnection in the wake of the erupting sigmoid and lie on the both sides of the center curved PIL (Figure 10(f)). The large quasi--circular ribbon was formed after the null--point reconnection and is located in the outer positive polarity regions (Figure 10(f)). However, we do not see the center compact ribbon in the observations. We believe that it may merge with the negative polarity parallel ribbon. Moreover, the circular ribbon is observed as an extended form of the parallel ribbon in the positive polarity. The observations show that both set of ribbons are formed after the sigmoid eruption. A detailed examination of the active region reveals fine thread--like structures (i.e., fine flux tubes) joining the center negative polarity with the circular ribbon (Figure 10(g)). Some parts of them are shown by the dashed yellow lines. We also find that the area surrounded by the circular ribbon is much brighter than the outer region in hotter AIA channels (Figures 10(b), (e) and (g)).

RHESSI observations are available between 12:50 UT and 13:10 UT, covering some part of the impulsive and peak phases of the main flare. Figure 11 shows the SDO/AIA 1600 \AA~images overplotted with the RHESSI X--ray contours from 12:51 to 12:55 UT. At 12:51 UT we observe RHESSI HXR sources in 25--50 (green) and 50--100 (blue) keV energy ranges that are co--spatial with the parallel ribbon kernels. The kernels are the high intensity regions within the two parallel ribbons. The SXR sources in 6--12 and 12--25 keV are located at the middle and are over the positive polarity ribbon in disk projection (Figure 11(a)). One minute later at 12:53 UT, the HXR sources remain in the same location on the same kernels, while the SXR sources move in the southwest direction. A westward shift of the western footpoint source has been also observed. The HXR sources remain more or less observed there from 12:54 to 12:55 UT, while there is a consistent upward motion of the SXR sources. HXR sources are usually formed at the footpoints, while SXR sources are usually coronal sources that originate from flare loops. The morphology and dynamics of RHESSI HXR and SXR sources match well with the standard model of solar flares. All these sources are believed to be formed after the reconnection between the surrounding arcades beneath the erupting sigmoid. During the decay phase a set of post flare loops are observed joining the two parallel ribbons (Figures 5(g)--(i)). The time sequence of the whole event is summarized in Table 1.

%------------------------------------------------------------------------------------------

\section{NLFFF Extrapolation Over the Active Region}
\label{}

In order to check the magnetic field configuration over the active region, we have performed a NLFFF extrapolation around the active region (Figure 12). The NLFFF extrapolation was carried out using the ``weighted optimization" \citep{Wheatland00,Wiegelmann04} method optimized for SDO/HMI magnetic field data \citep{Wiegelmann10,Wiegelmann12}, based on a preflare HMI vector magnetogram at 12:20 UT. A preprocessing procedure \citep{Wiegelmann06} was applied to make the photospheric boundary suit the force--free condition. Figure 12(a) shows the radial component of the SDO/HMI magnetic field at $\approx$12:20 UT. The negative and positive polarity regions are marked as N1, N2 and P1, P2, respectively. The extrapolated field lines over the magnetic field are shown in Figure 12(b). The white box in Figures 12(a) and 12(b) is the region of tether--cutting type of reconnection. We see several sheared field lines connecting the regions N2, P2 and N1, P1. These results match very well with the observations. The area within the box is the region of initial compact brightenings observed during the preflare and the main flare phases (c.f., Figures 3(a)--(c) and 5(a)--(b)). This brightening is believed due to the reconnection between the sheared field lines at the PIL between N1 and P2. Figure 12(b) also shows the large--scale loops that fan out from the central negative polarity region and connect to the surrounding weaker positive polarity region. The extrapolated large--scale field structure matches quit well with our observations (c.f., Figure 10). Many of the observed fine thread--like structures in SDO/AIA 94 \AA~match well with the extrapolated lines (c.f., Figures 10(g) and 12(b)). Overall, the extrapolation result indicates sheared and sigmoidal field lines at the flaring core region, enveloped by large--scale magnetic fields with a fan--like structure. However, we did not find a coronal null point in the NLFFF model.
%------------------------------------------------------------------------------------------

\section{Results and Discussions}
\label{}

In this paper, we present the multiwavelength observations and interpretation of a sigmoid eruption and its possible role in triggering a solar flare with parallel ribbons and a quasi--circular ribbon in a large--scale fan--spine type magnetic configuration. NLFFF extrapolation analysis has been also carried out to understand the surrounding magnetic field structure over the active region. Two different stages of reconnections have been also discussed. We also discuss the preflare phase and the nearby jet activity and their linkage to the main flare phase. The key findings of this study are the followings.

1. Several observational signatures and NLFFF results suggest the existence of a large--scale fan--spine type magnetic configuration with a large sigmoid embedded in a section of the fan dome.

2. Our observations show two main stages of evolution of the whole event, i.e., preflare, and main flare including a nearby jet activity.

3. During the preflare phase, we observed the formation and appearance of the sigmoid underneath the southern section of the fan--dome structure.

4. The internal tether--cutting type reconnection below the pre--existing sigmoid may be responsible for the sigmoid eruption.

5. Our analysis suggests that the erupting sigmoid trigger two stages of reconnection. The first stage is the reconnection between the legs of surrounding arcade field lines at the wake of the erupting sigmoid which could produce the two parallel ribbons. The second stage involves the null--point reconnection, leading to the formation of the large--scale circular ribbon.

The fan--spine type magnetic configuration with an underneath sigmoid has been observed to produce quasi--circular ribbon solar flares \citep{Masson09,Su09,Reid12,Wang12,Sun13,Jiang13,Jiang14}. In our study, we found various observational signatures that support the existence of this kind of magnetic configuration. The extrapolated NLFFF lines over the active region are likely to be consistence with the observations. Formation of the typical quasi--circular flare ribbon provides the observational evidence of the presence of the fan--spine type magnetic configuration \citep{Masson09,Wang12,Reid12,Dai13,Sun13,Jiang13,Jiang14}. In our case, we also observed a typical large--scale circular ribbon that does not expand much, signifying the existence of the fan--spine type configuration (see Figures 9, 10 and the associated animations). However, we did not observe a compact central ribbon that was usually detected at the footpoint of the inner spine \citep[e.g.,][]{Masson09,Wang12}. We conjecture that the central ribbon may merge with the negative polarity parallel ribbon. It is also evident from the analysis that the circular ribbon lies at the positive polarity region around the center negative polarity region (Figure 10(f)). The NLFFF extrapolation result also shows that fan--like field lines connect the outer circular--ribbon regions with positive polarity to the central negative polarity region (c.f., Figures 12(b)). Also, during the time of appearance of the circular ribbon in hotter corona AIA channels, we found a high intensity area between the circular ribbon and the main flare site compared to the outer quiet area (Figures 10(b) and (e)). This also provides an observational signature of the existence of a closed dome--like structure which is visible when hot plasma filled the fan lines. The NLFFF extrapolation also reproduces large--scale field lines that appears like a dome (c.f., Figures 12(b)), which is very similar to the observations. We also found the observational evidence for the existence of sigmoid in the hotter temperature (Figure 3(e)), which is believed to be located underneath the southern section of the fan dome. Previous evidence for the existence of such kind of topology in large--scale has rarely been obtained \citep{Jiang13,Jiang14}. However, in our NLFFF model analysis a coronal null related to the large scale circular ribbon may not exist.

The whole event can be understood with the following two different phases. The preflare phase contains the dynamics of the formation and appearance of large sigmoid underneath the large fan dome structure (Section 3.2 and animation associated with Figure 3). The formation of the sigmoid can be explained via the tether--cutting reconnection between the low--lying sheared field lines (see the schematic representation in Figure 4 and \cite{Moore01}). The formation process starts with the interaction and reconnection between the two sets of sheared field lines at the PIL (Figures 3(a) and 4(a)). The existence of RHESSI X--ray sources near the merging area shows the evidence of magnetic reconnection (Figures 3(c) and 4(b)). The two sets of sheared lines reconnected and formed the sigmoid (Figures 3(e) and 4(c)). Recently, \cite{Cheng15} also studied the formation of the same sigmoid using {\it Interface Region Imaging Spectrograph} (IRIS) spectroscopic and AIA imaging observations and interpreted its formation due to the flux cancellation process. We believe that the coronal tether--cutting reconnection is more appropriate to explain the formation of the sigmoid in the preflare phase. Similar interpretations of the formation of the sigmoid have been reached in few recent studies \citep{Liu13,Jiang14,Joshi14a}. \cite{Jiang14} found a similar observation of the formation of the sigmoid underneath the fan dome and interpreted it as the reconnection between the sheared field lines. We also observed the appearance of the sigmoid after the hot plasma filled the field lines (see Figure 3). Similar hot flux ropes signatures have been also observed in the past studies after some reconnection--driven heating of flux rope plasma \citep{Li13,Joshi14a}. Apart from this, it may be also possible that some part of the flux rope already exists there within the fan dome and during the preflare reconnection, where some new flux may be added to the pre--existing flux rope. The sigmoid that appeared may be a part of the existing flux rope.

The triggering of the sigmoid eruption, the main flare reconnections and the formation of different sets of flare ribbons may be summarized using a simple schematic picture shown in Figure 13. The initial configuration is believed to be a fan--spine type with inner (blue) and outer (red) fan lines with a null point (Figure 13(a)). The pre--existing sigmoid is represented by the thick black line. The sigmoid eruption and the formation of parallel ribbons can be interpreted using the tether--cutting scenario \citep[c.f., ][]{Moore01,Liu07}. First, the main flare starts with the tether--cutting reconnection between the low--lying sheared field lines (shown by green color in Figure 13(a). The compact brightening in the junction of the low--lying sheared lines in the middle is strong evidence of the tether--cutting reconnection (Figure 5(a)). This can be interpreted as the first stage of the tether--cutting reconnection model. The reconnection region is shown by the white star in Figure 13(a). This leads to the formation of large sigmoidal field lines, which may merge with the pre--existing field of the sigmoid (thick green/black line in Figure 13(b)). This low--lying reconnection is believed to be responsible for the eruption of the sigmoid. This mechanism of sigmoid eruption is consistent with the results of \cite{Liu07} and \cite{Savcheva14}. We also observed a jet--like activity near this tether--cutting type reconnection region. We suspect that this jet--like activity may play some roles in triggering the sigmoid eruption by disturbing the surrounding magnetic field of the sigmoid (Figures 7, 13(a) and the associated animation).

It is believed that the second stage of reconnection occurred underneath the sigmoid eruption between the overlying arcade field lines triggers the first stage of reconnection during eruption. This results in the formation of the parallel ribbons when the accelerated electrons hit the low solar atmosphere (Figures 9(d) and 13(b)). This scenario can be well supported by the standard solar flare model also known as the ``CSHKP" model \citep{Carmichael64,Sturrock66,Hirayama74,Koop76} as well as the second stage of the tether--cutting model \citep[c.f., Figure 1 of][]{Moore01}. Several pieces of observation have been observed that are consistent with these models, including (1) the temporal correlation between the sigmoid eruption and the GOES flux enhancement (Figure 7), (2) the formation of parallel ribbons simultaneously with the eruption of the sigmoid (Figures 7(b) and 9), and (3) the observations of the HXR sources at the brightest kernels of the ribbons and their apparent separation motion (Figure 11). All these observational signatures are quite strong to support the fact that the second stage flare reconnection is in between the surrounding arcades in the wake of a sigmoid eruption.

Later on, the erupting sigmoid also triggered a null--point reconnection between the inner (blue) and outer (red) fan field lines at the null point (Figure 13(b)). This is the third stage of reconnection. The null--point reconnection accelerated the electrons towards the fan dome footpoints and produced the circular ribbon (Figures 9, 10 and 13(b)). Moreover, the circular ribbon is seen as an extension of the parallel ribbon in the positive polarity region (Figure 9(d)--(f)). We conjecture that the center ribbon may coincide with the negative polarity parallel ribbon. Till now only one flare event on 2011 September 6 shows the existence of two different sets of flare ribbons \citep{Jiang13,Jiang14}. This kind of null--point reconnection and the associated circular or quasi--circular ribbon flares may be triggered either by the shearing motion of the fan field lines of fan--dome structure \citep{Vemareddy14b} or due to the eruption of the sigmoid underneath the fan dome structure \citep{Sun13,Dai13,Jiang13,Jiang14}. \cite{Jiang13,Jiang14} found that the underlying sigmoid was in the domain of torus instability. The magnetic reconnection within the low--lying sigmoid triggered its initial expulsion, which later triggered the breakout reconnection at the magnetic null. In our case, we also found that the erupting sigmoid triggered by the low--lying tether--cutting type reconnection caused the null--point reconnection later on. Indeed, we also found that the sigmoid was rising about 11 minutes before the first appearance of the circular ribbon, which may provide another strong piece of evidence that the rising sigmoid was responsible to trigger null point reconnection (Figure 9(d). However, our NLFFF model results do not show the location of the null point, but we strongly believe the existence of null point higher in corona. We also did not find signatures of remote brightenings. One possibility is that the intensity of the remote brightening may be too weak to be detected. The apparent counterclockwise brightening enhancement (i.e., starting from west and expanding to the east) along the circular ribbon has been observed (see the animation associated with Figure 9). This suggests that the reconnection at the null may be the slipping--reconnection type. Similar kinds of counterclockwise brightening motions have been also observed in several studies but only for the smaller--scale flares \citep{Masson09,Wang12,Sun13}. For the first time we are suggesting this type of two stage reconnection and the formation of two different sets of ribbons in great detail.

No exact mechanism has been proposed for the triggering of flare producing this kind of two set of flare ribbons (i.e., parallel and circular) and its association with the sigmoid eruptions. However, most of the results reveal that circular ribbon flares may be due to the null point type reconnection in a typical fan--spine configuration. More observational and simulation studies are required to understand these kind of flares in small as well as large--scale. In the future, we will try to work on similar more such events in order to understand the exact mechanism behind the triggering of flare reconnection which produces two different sets of flare ribbons i.e., parallel and large--scale circular ribbon flares via sigmoid eruption.

%------------------------------------------------------------------------------------------

\acknowledgments
The authors thank the referee for his/her valuable comments and suggestions. We thank SDO/AIA, SDO/HMI, NSO--GONG, GOES and RHESSI teams for providing their data for the present study. This work was supported by the BK21 plus program through the National Research Foundation (NRF) funded by the Ministry of Education of Korea, Basic Science Research Program through the NRF funded by the Ministry of Education (NRF--2013R1A1A2012763), NRF of Korea Grant funded by the Korean Government (NRF--2013M1A3A3A02042232), and the Korea Meteorological Administration/National Meteorological Satellite Center. NCJ thanks School of Space Research, Kyung Hee University for providing the postdoctoral grant. CL and HW are supported by NSF, under grants AGS--1348513 and AGS--1408703, and by NASA under grants NNX13AG13G, NNX13AF76G and NNX14AC12G.    

%------------------------------------------------------------------------------------------
%\bibliography{reference.bib}
%\bibliographystyle{apj} % style aa.bst
%\bibliography{reference} % your references Yourfile.bib

%------------------------------------------------------------------------------------------
%----------------------------------------------------------------------------
\clearpage
\begin{figure}
\vspace*{-3cm}
\centerline{
	\hspace*{0.0\textwidth}
	\includegraphics[width=2.5\textwidth,clip=]{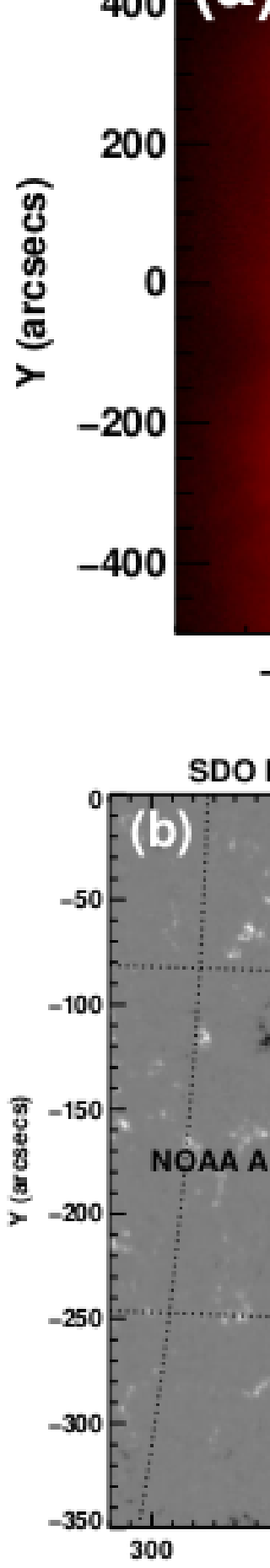}
	}
\vspace*{-5.5cm}
\caption{(a) SDO/AIA 304 \AA~image showing the location of the activity region under study on the Sun marked by the black box. (b) and (c) display SDO/HMI line--of--sight magnetogram and SDO/AIA continuum image respectively corresponding to the field of view of the black box in the panel (a). The area covered by the white box is used for calculating intensity profiles at different wavelengths (see Figure 2(b)).}
\label{}
\end{figure}
%------------------------------------------------------------------------------------
\clearpage
\begin{figure}
\centerline{
	\hspace*{0.0\textwidth}
	\includegraphics[width=1.8\textwidth,clip=]{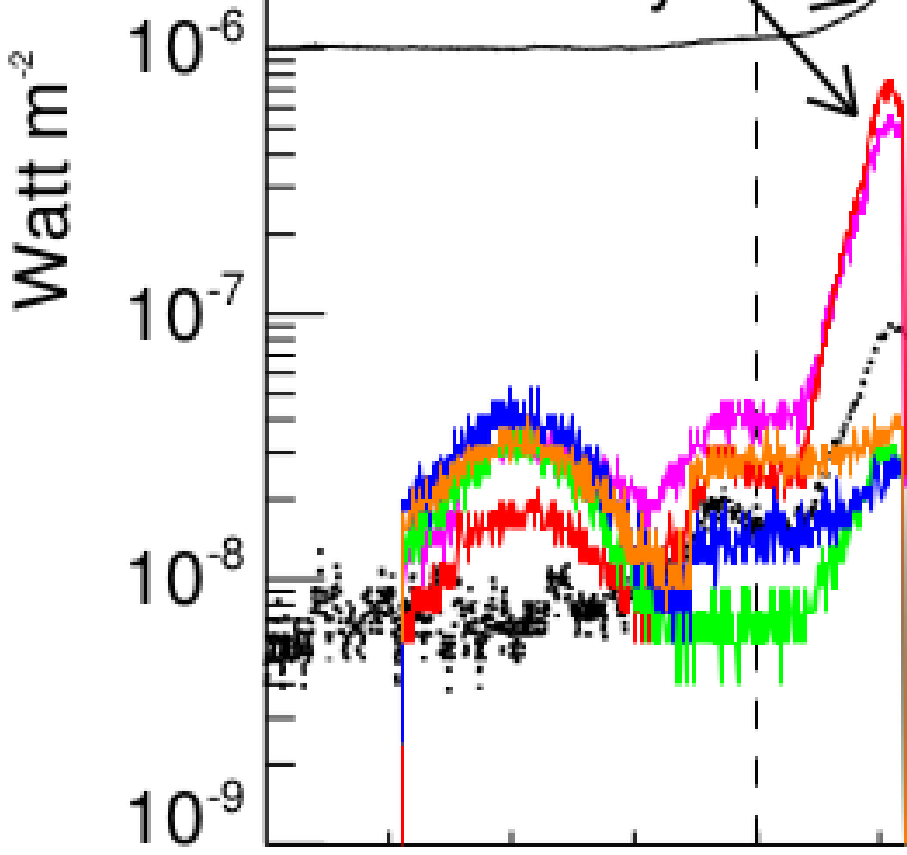}
	}
\vspace*{-1.5cm}
\caption{(a) GOES (black) and RHESSI (color) X--ray time profiles between 11:00 UT to 15:00 UT on 2014 April 18. The profiles show three different phases i.e., the preflare, the nearby jet activity and the main M7.3 flare. Region between the two vertical dashed black lines represent the preflare activity phase. Areas between the two dashed red and blue lines show the nearby jet activity and the main flare phases, respectively. (b) Normalized intensity profiles during 11:35 UT to 12:46 UT for SDO/AIA 131, 94, 171 and 1600 \AA~wavelength channels. The area used for these intensity measurements is represented by the white box in Figure 1(a).}
\label{}
\end{figure}
%------------------------------------------------------------------------------------
\clearpage
\begin{figure}
\centerline{
	\hspace*{0.0\textwidth}
	\includegraphics[width=1.4\textwidth,clip=]{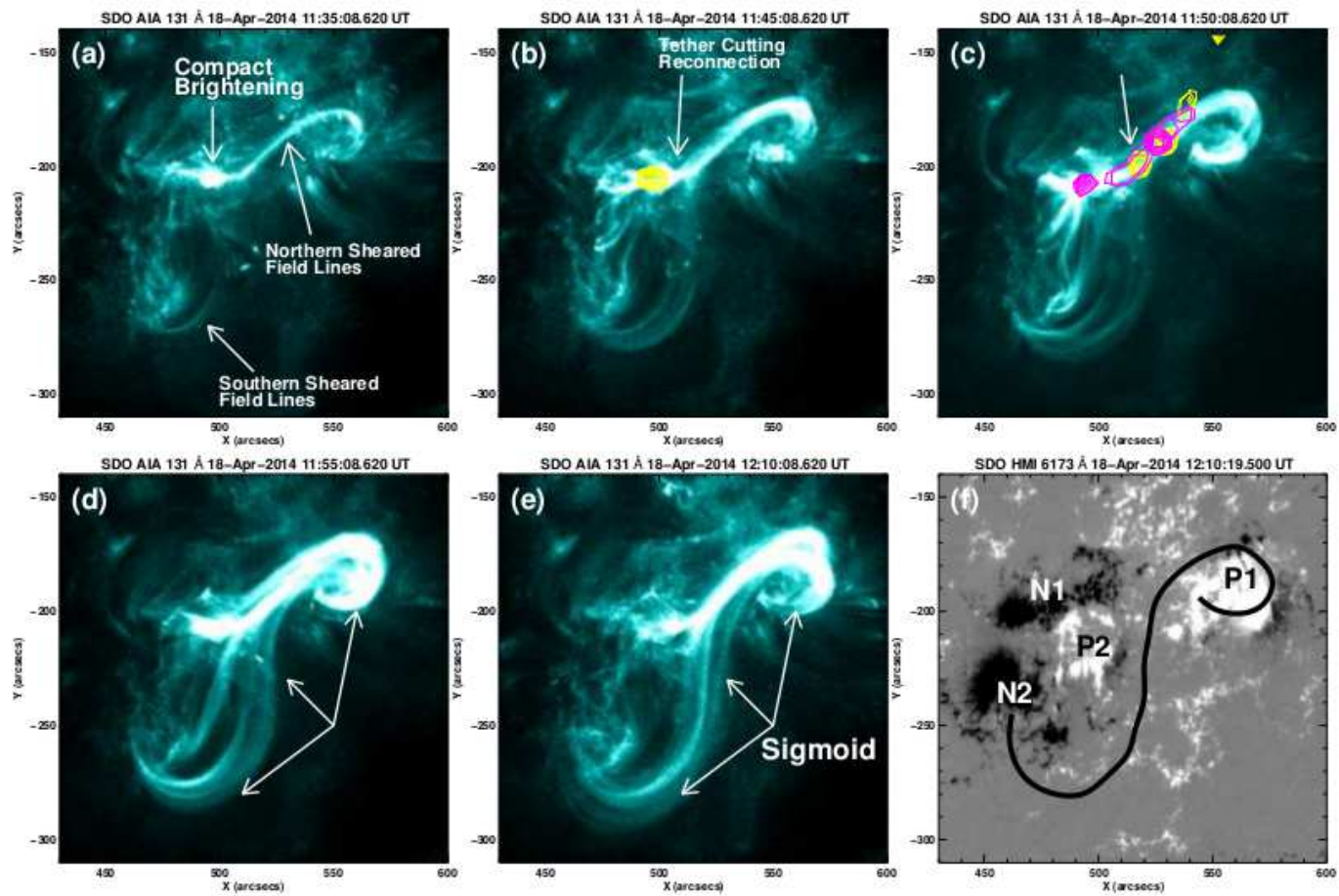}
	}
\vspace*{-2.5cm}
\caption{((a)--(e)) Selected SDO/AIA 131 \AA~images showing the formation and appearance of the sigmoid. The pink and yellow contours are the RHESSI X--ray contours at 3--6 and 6--12 keV energy bands. The contour levels are 35\%, 40\%, 50\%, 60\%, 70\%, 80\%, 90\%, and 95\% of the peak intensity. The integration time is 20 s. (f) The SDO/HMI line--of--sight magnetogram at 12:10:19 UT. We overplotted an drawing of approximate sigmoid axis with black color using Figure 3(e). Here 'N' and 'P' letters represent the negative and positive polarities, respectively.} 
\label{}
\end{figure}
%------------------------------------------------------------------------------------
\clearpage
\begin{figure}
\centerline{
	\hspace*{0.0\textwidth}
	\includegraphics[width=1.4\textwidth,clip=]{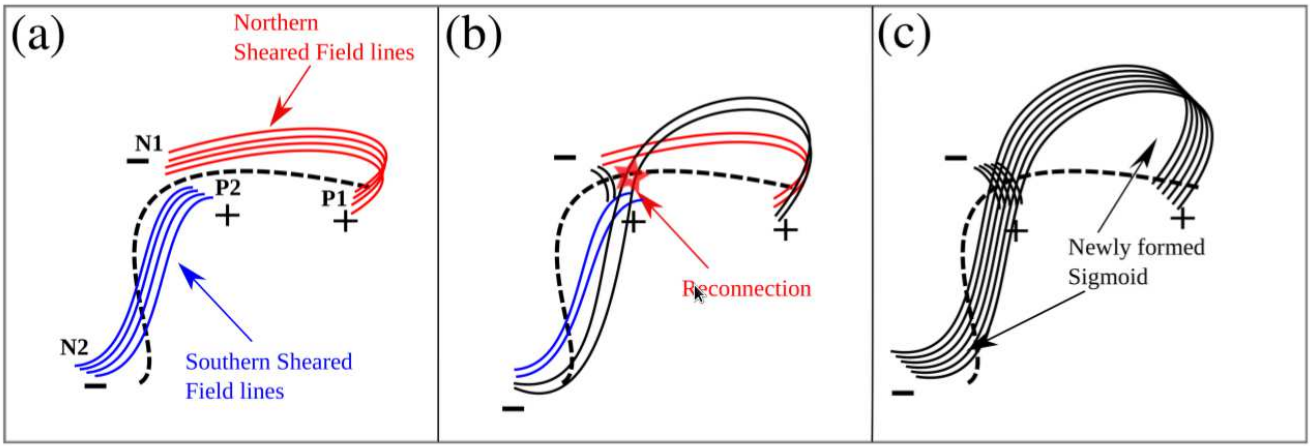}
	}
\vspace*{-4cm}
\caption{Schematic of the preflare phase. The red and blue lines show the northward and southward sheared field lines. The black lines represent the magnetic lines formed after the reconnection. The dashed line shows the quasi--circular polarity inversion line.}
\label{}
\end{figure}
%------------------------------------------------------------------------------------
\clearpage
\begin{figure}
\centerline{
	\hspace*{0.0\textwidth}
	\includegraphics[width=1.6\textwidth,clip=]{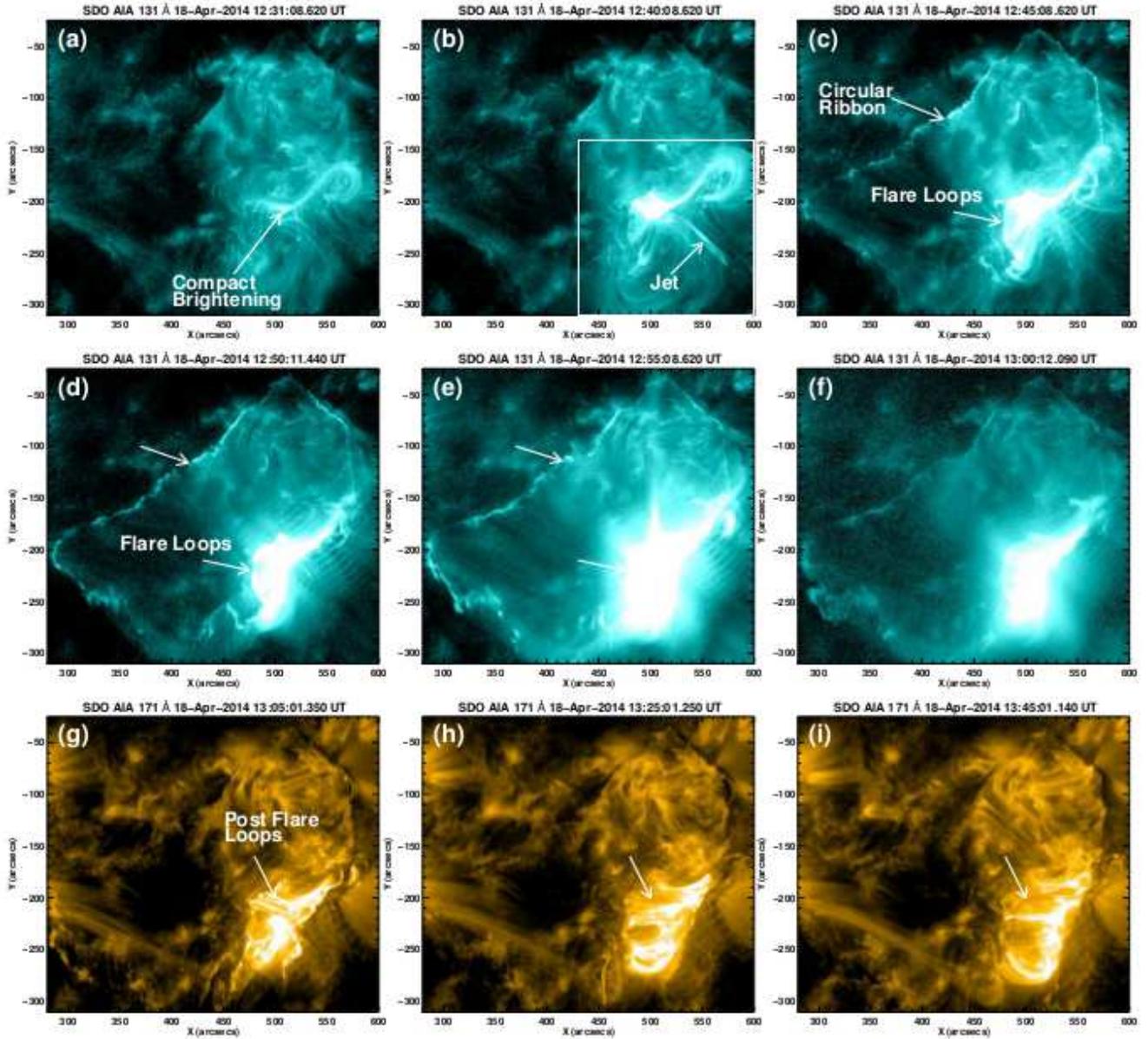}
	}
\vspace*{-1.8cm}
\caption{((a)--(f)) Selected SDO/AIA 131 \AA~images showing the evolution of the main flare phase and the nearby jet. The decay phase of the main flare is shown by the SDO/AIA 171 \AA~images ((g)--(i)).} 
\label{}
\end{figure}
%------------------------------------------------------------------------------------
\clearpage
\begin{figure}
\centerline{
	\hspace*{0.0\textwidth}
	\includegraphics[width=1.4\textwidth,clip=]{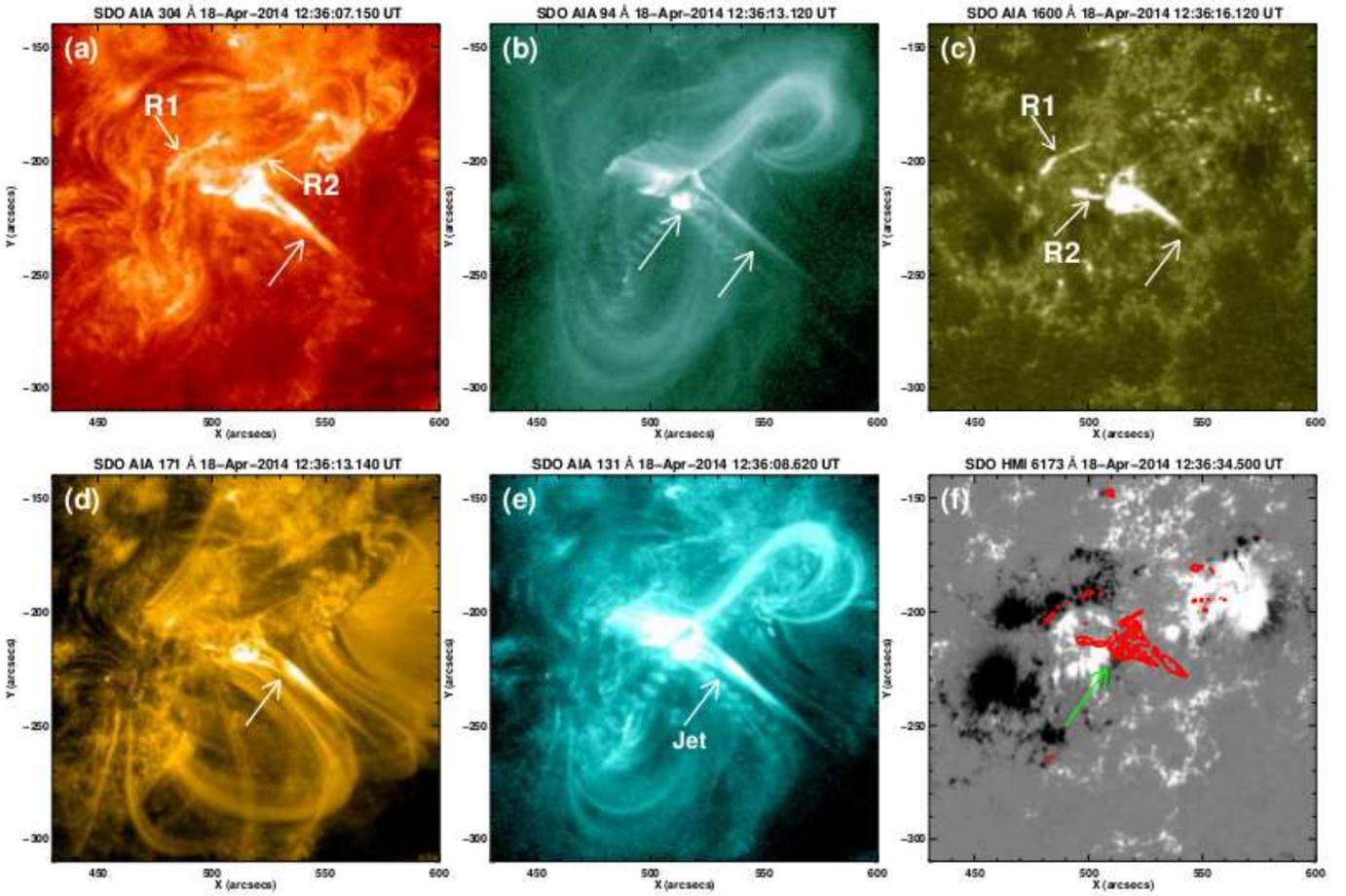}
	}
\vspace*{-2.5cm}
\caption{((a)--(e)) SDO/AIA images in 304, 94, 1600, 171, 131 \AA~channels at $\approx$12:36 UT on 2014 April 18 showing the nearby jet. (f) The SDO/HMI line--of--sight magnetogram overplotted by the SDO/AIA 304 \AA~intensity contours in red color. The contour levels are 8\%, 10\%, 20\%, 30\%, 40\%, 50\%, 60\%, 70\%, 80\%, and 90\% of the peak intensity. The green arrow represents the footpoint of the Jet.}
\label{}
\end{figure}
%------------------------------------------------------------------------------------
\clearpage
\begin{figure}
\centerline{
	\hspace*{0.0\textwidth}
	\includegraphics[width=1.6\textwidth,clip=]{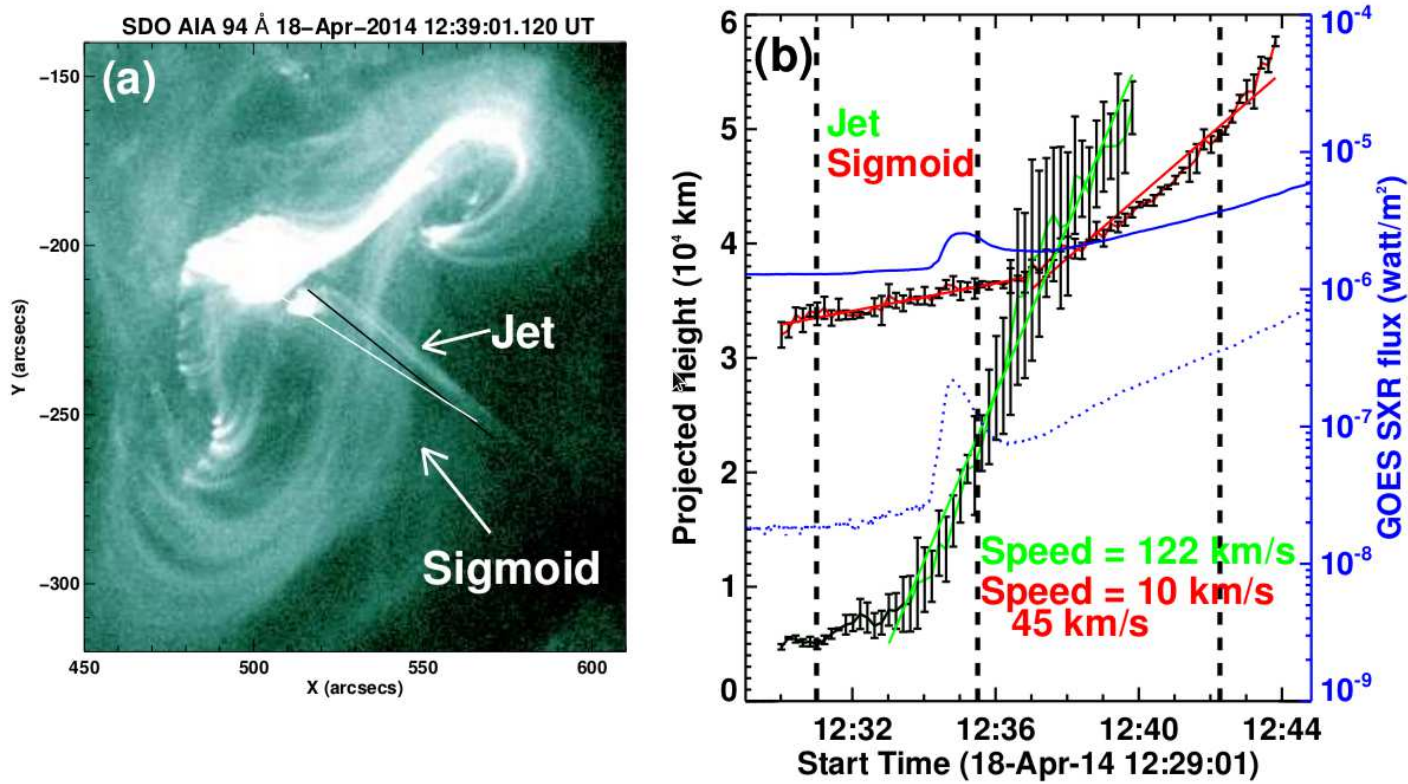}
	}
\vspace*{-3.0cm}
\caption{(a) SDO/AIA 94 \AA~image at $\approx$12:39 UT. Overplotted white and black lines represent the approximate trajectories along with the projected height--time measurements have been estimated for the jet ejection and sigmoid eruption, respectively. (b) Projected height--time profiles of the jet and the sigmoid are shown with green and red colors, respectively. The error bars are the standard deviations estimated after the three repeated measurements. The speeds are calculated by the linear fits to the height--time data points. Solid and dotted blue curves represent the GOES X--ray flux profiles in 1--8 and 0.5--4 \AA~channels respectively. The dashed lines from left to right show the start time of the main flare (12:31 UT), first appearance time of the parallel ribbons (12:35:28 UT) and the first appearance time of the circular ribbon (12:42:16 UT), respectively.} 
\label{}
\end{figure}
%------------------------------------------------------------------------------------
\clearpage
\begin{figure}
\centerline{
	\hspace*{0.0\textwidth}
	\includegraphics[width=1.2\textwidth,clip=]{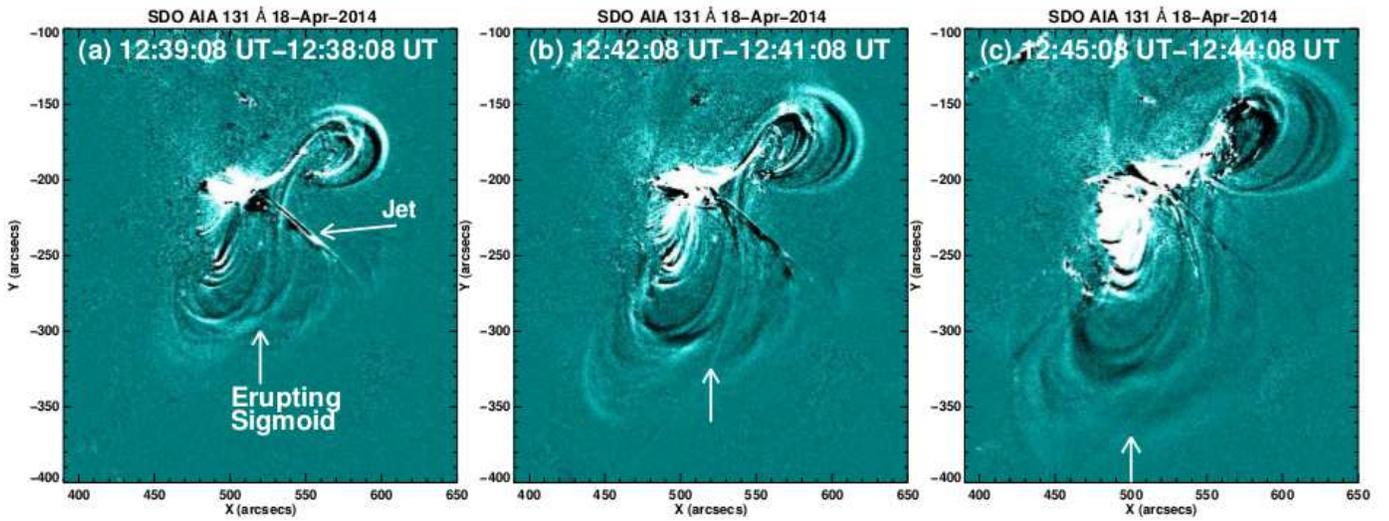}
	}
\vspace*{-4cm}
\caption{((a)--(c)) SDO/AIA 131 \AA~ running difference images showing the eruption of the sigmoid between $\approx$12:39 UT to $\approx$12:45 UT on 2014 April 18. Vertical arrows represent the approximate top most part of the erupting sigmoid.} 
\label{}
\end{figure}
%------------------------------------------------------------------------------------
\clearpage
\begin{figure}
\centerline{
	\hspace*{0.0\textwidth}
	\includegraphics[width=1.2\textwidth,clip=]{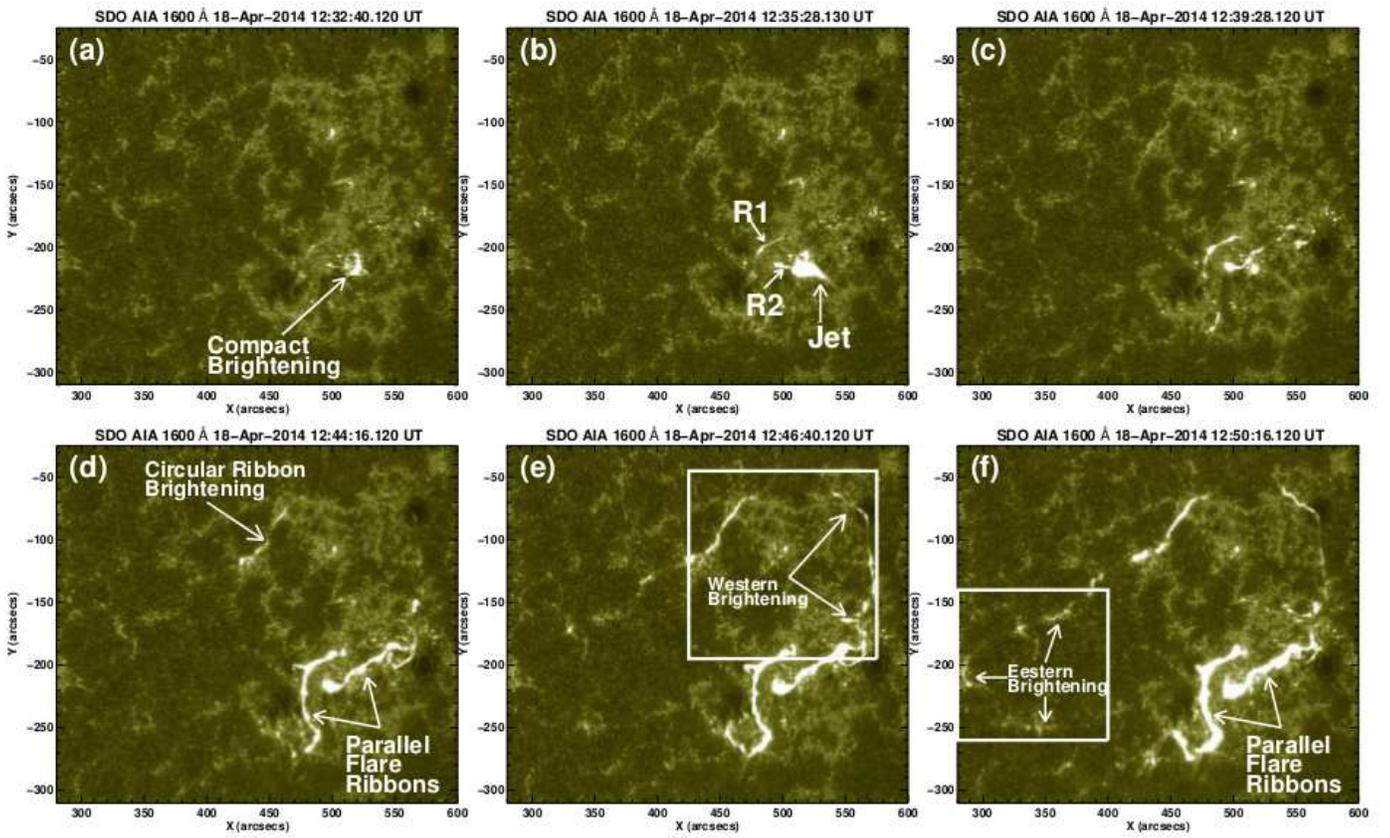}
	}
\vspace*{-2cm}
\caption{SDO/AIA 1600 \AA~images showing the formation and evolution of the parallel ribbons and the large--scale quasi--circular ribbon.} 
\label{}
\end{figure}
%------------------------------------------------------------------------------------
\clearpage
\begin{figure}
\vspace*{-3cm}
\centerline{
	\hspace*{0.0\textwidth}
	\includegraphics[width=2.5\textwidth,clip=]{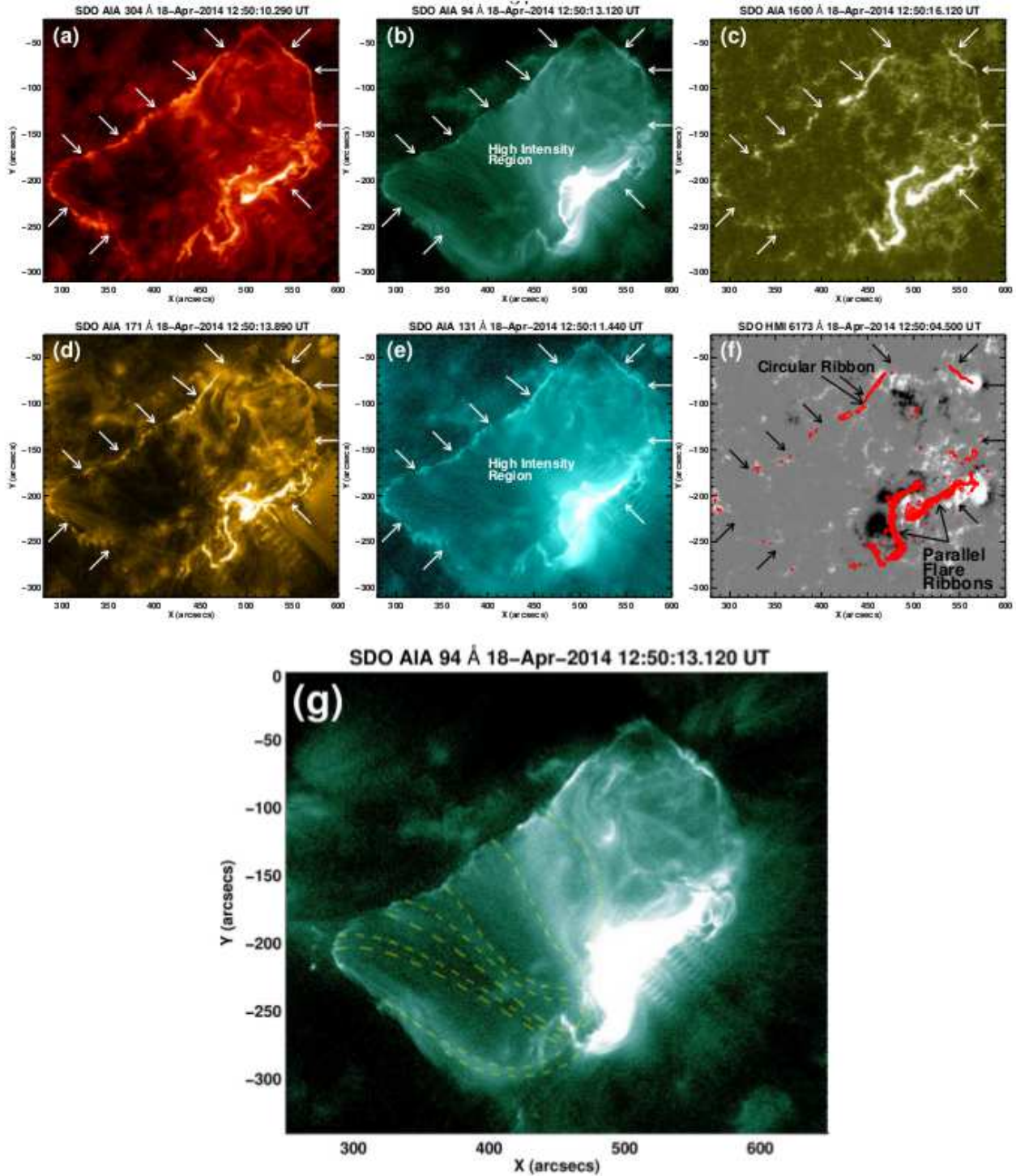}
	}
\vspace*{-3.0cm}
\caption{((a)--(e)) SDO/AIA images in 304, 94, 1600, 171, 131 \AA~channels at $\approx$12:50 UT on 2014 April 18 showing the parallel ribbons and the circular ribbon (with white arrows). (f) The SDO/HMI line--of--sight magnetogram overplotted by the SDO/AIA 1600 \AA~intensity contours in red color. The contour levels are 5\%, 10\%, 20\%, 30\%, 40\%, 50\%, 60\%, 70\%, 80\%, and 90\% of peak intensity. (g) SDO/AIA 94 \AA~image at $\approx$12:50 UT. Few visible fine threads (i.e., flux tubes) joining the center negative to the circular positive polarity ribbon are tracked and shown by the dashed yellow lines.} 
\label{}
\end{figure}
%------------------------------------------------------------------------------------
\clearpage
\begin{figure}
\centerline{
	\hspace*{0.0\textwidth}
	\includegraphics[width=1.4\textwidth,clip=]{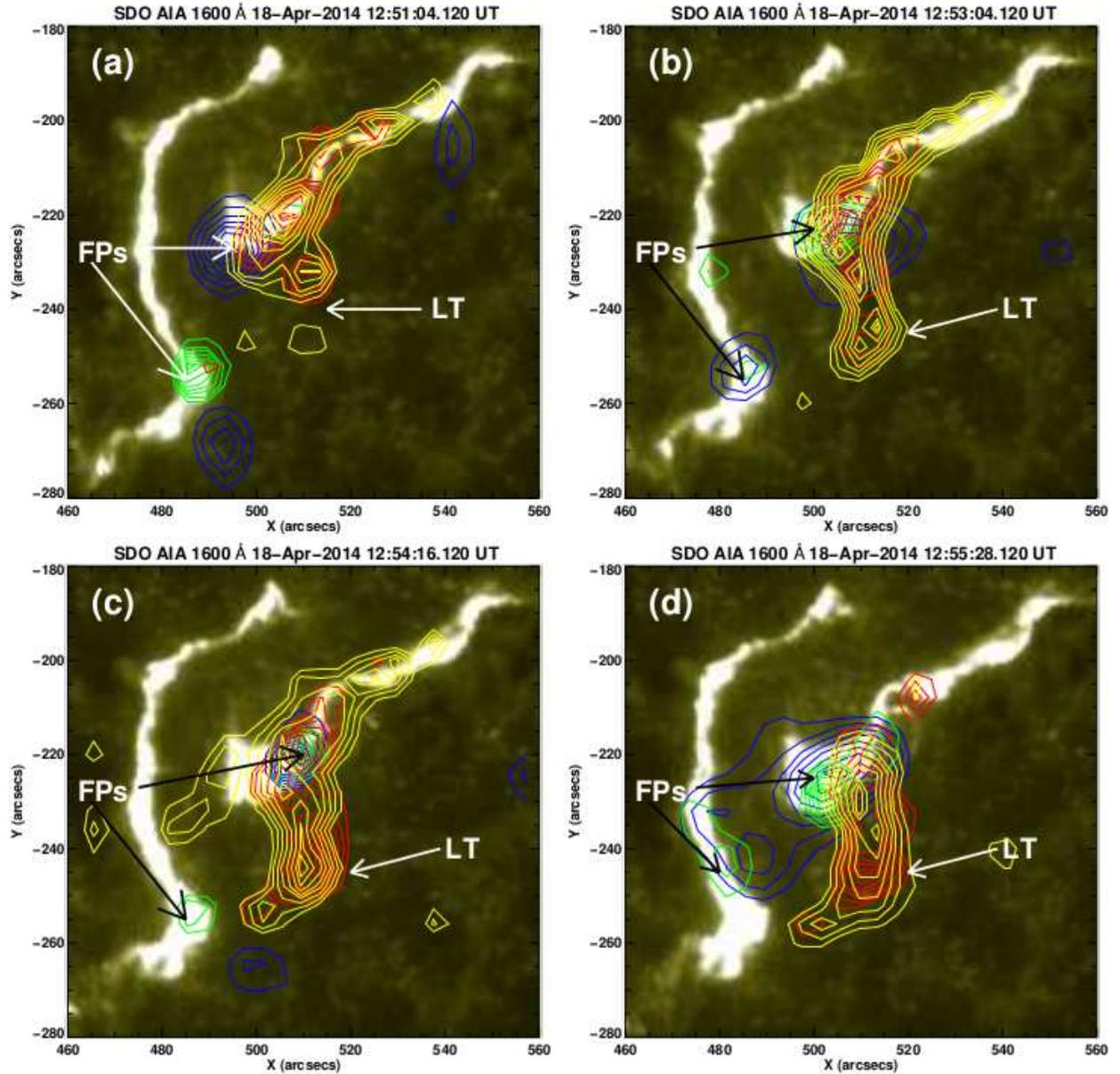}
	}
\vspace*{-0.5cm}
\caption{SDO/AIA 1600 \AA~images overplotted with the RHESSI X--ray contours, showing the evolution of coronal and footpoint sources. The yellow, red, green and blue contours are the RHESSI X--ray contours at 6--12, 12--25, 25--50 and 50--100 keV energy bands. The contour levels are 30\%, 40\%, 50\%, 60\%, 70\%, 80\%, 90\%, and 95\% of peak intensity. The integration time is 20 s. Footpoints (FPs) and approximate top of flare loops (LT) are also marked.}
\label{}
\end{figure}
%------------------------------------------------------------------------------------
\clearpage
\begin{figure}
\centerline{
	\hspace*{0.0\textwidth}
	\includegraphics[width=1.5\textwidth,clip=]{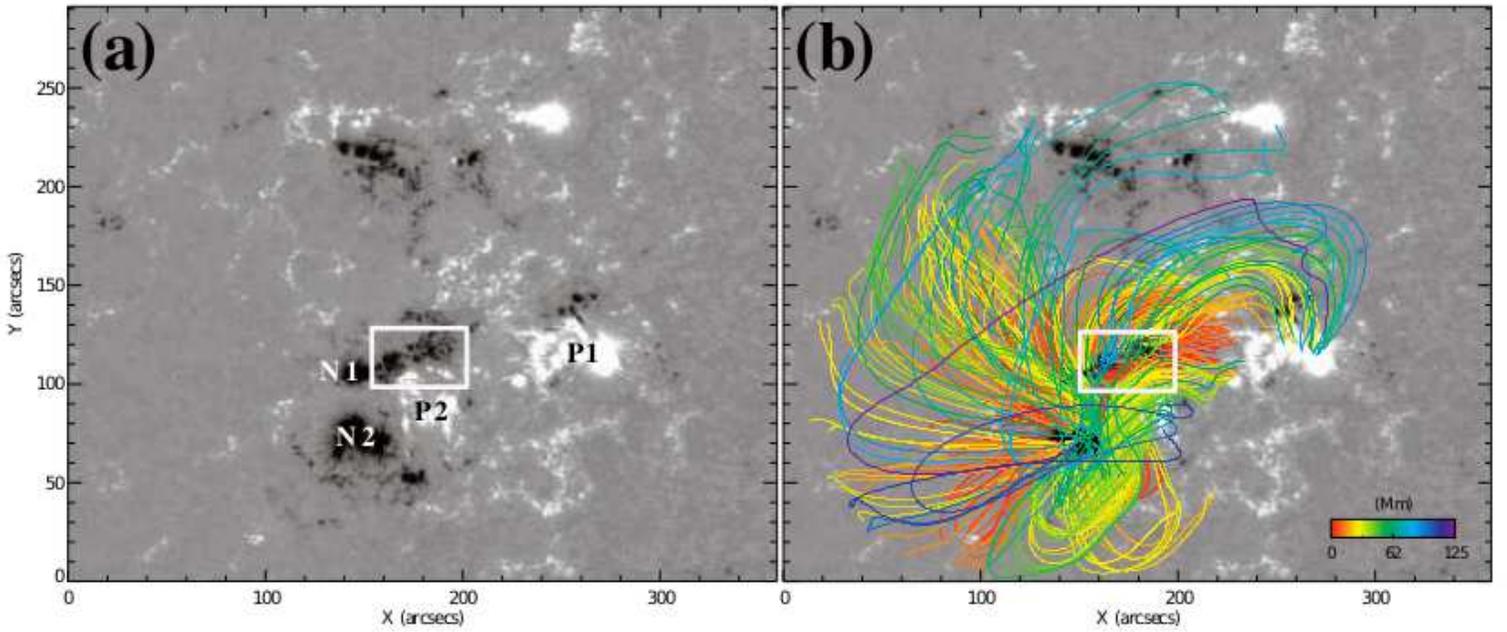}
	}
\vspace*{-4.5cm}
\caption{(a) SDO/HMI vertical field remapped using Lambert equal area projection at $\sim$12:20 UT on 2014 April 18. (b) Same as (a) but overplotted with selected NLFFF lines (colored according to the maximum height), showing the sigmoidal and the overarching fan--like fields.}
\label{}
\end{figure}
%------------------------------------------------------------------------------------
\clearpage
\begin{figure}
\vspace*{-3cm}
\centerline{
	\hspace*{0.0\textwidth}
	\includegraphics[width=2.5\textwidth,clip=]{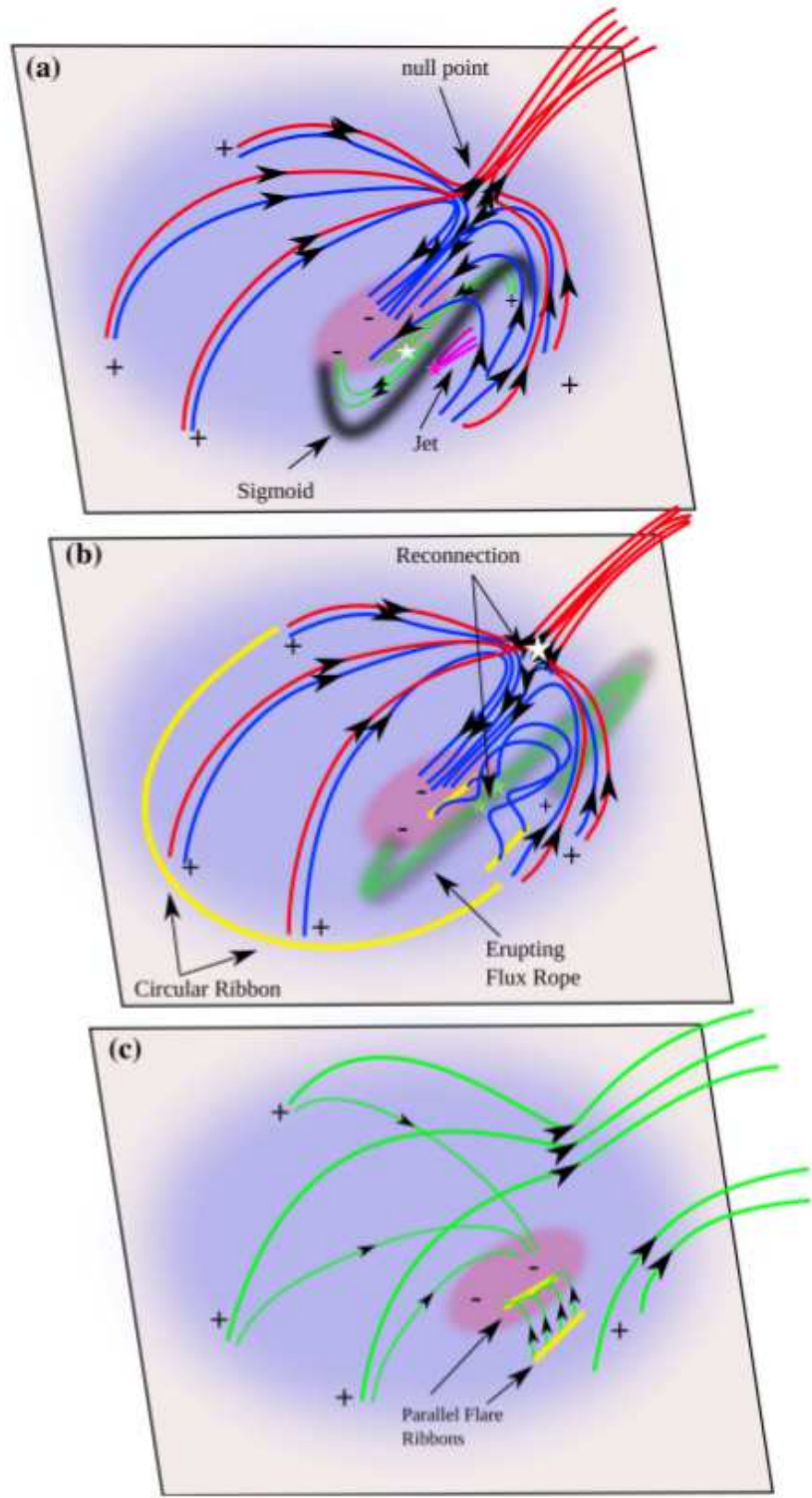}
	}
\vspace*{-3.0cm}
\caption{Schematic representation showing the magnetic configuration and the triggering of the main M7.3 flare. Blue/Red lines show the inner/outer field lines of the fan dome. Green lines in panel (a) represent the low--laying sheared field lines. The jet is shown by the pink color in panel (a). The pre--existing sigmoid is represented by the black line in panel (a). Newly formed sigmoid field lines are shown by the green color in panel (b). The parallel ribbons and the circular ribbon are shown by the yellow color in panels (b) and (c). The changed magnetic configuration is shown in panel (c).}
\label{}
\end{figure}
%------------------------------------------------------------------------------------------
%% Table
%
\clearpage
\begin{table}
\small
%\begin{sidewaystable}
\caption{Time Sequence of the Event.}
%\resizebox{\textwidth}{!}{%
\label{table1}
\begin{tabular}{llp{9cm}}
\hline
Event Phase & Time & Observed activities\\
\hline
%\rotate
%\multicolumn{3}{c}{<>}
Pre Flare 			  &$\sim$11:35 UT   & Compact brightening observed near the junction\\
($\sim$11:35 UT--$\sim$12:10 UT)& & of the northward and southward sheared lines.\\
				  &$\sim$10:35--$\sim$12:10 UT & Sigmoid formation and appearance via tether--cutting reconnection.\\
		   		  &$\sim$12:10 UT & Full appearance of the sigmoid structure.\\ 
\hline
Main Flare 			  &$\sim$12:31 UT & (a) Start time of the main flare M7.3,\\
($\sim$12:31 UT--$\sim$14:40 UT)  & & (b) compact brightening started at the junction of sheared field lines, (c) start of the initial slow eruption of the sigmoid.\\
				  &$\sim$12:31 UT--$\sim$12:37 UT & Initial slow eruption phase of the sigmoid ($\rm\sim~10~km~s^{-1}$).\\
				  &$\sim$12:34 UT--$\sim$12:36 UT & Duration of nearby jet activity.\\
	   			  &$\sim$12:35 UT & First appearance of parallel ribbons.\\
				  &$\sim$12:37 UT--$\sim$12:44 UT & Acceleration phase of the sigmoid eruption ($\rm\sim~45~km~s^{-1}$).\\
	   			  &$\sim$12:42 UT & First appearance of the middle part of the circular ribbon.\\
				  &$\sim$12:44 UT & Well--developed parallel ribbons observed on both sides of center circular polarity inversion line.\\
				  &$\sim$12:44 UT--$\sim$13:30 UT & Parallel ribbons separation.\\
	   			  &$\sim$12:45 UT & Brightness in the western part of the circular ribbon.\\
	   			  &$\sim$12:46 UT--$\sim$12:50 UT & Brightness moves from west to east along the circular way.\\
	   			  &$\sim$12:50 UT & Well--developed circular ribbon observed.\\
	   			  &$\sim$12:51 UT--$\sim$12:55 UT & RHESSI coronal and footpoint X--ray sources observed that are co--spatial with the bright kernels of the parallel ribbons.\\
	   			  &$\sim$13:03 UT & Peak time of the main flare.\\
	   			  &$\sim$13:45 UT & Well--developed post flare loops are observed joining the parallel ribbons.\\
\hline
\end{tabular}
%\end{sidewaystable}
\end{table}
%------------------------------------------------------------------------------------
\end {document}